\documentclass[a4paper, journal]{IEEEtran}
\pdfoutput=1
\PassOptionsToPackage{bookmarks={false}}{hyperref}
\usepackage[pdftex]{graphicx}
\usepackage{multirow}
\usepackage{array,arydshln}
\usepackage{amsmath}

\usepackage{amsmath,amssymb}
\usepackage[tight,footnotesize]{subfigure}
\usepackage[linesnumbered,ruled,vlined,]{algorithm2e}
\usepackage{cite}
\DeclareMathOperator{\E}{E}

\begin{document}

\title{FEC-Assisted Parallel Decoding of \\Polar Coded Frames: Design Considerations}

\author{Mohammad Sadegh Mohammadi,
	Eryk Dutkiewicz,~and
    Qi Zhang    
\thanks{Mohammad Sadegh Mohammadi is with Aarhus University, Denmark and Macquarie University, Australia (email: msmo@eng.au.dk).}
\thanks{Eryk Dutkiewicz is with Macquarie University, Australia (email: eryk.dutkiewicz@mq.edu.au). }
\thanks{Qi Zhang is with Aarhus University, Denmark (email: qz@eng.au.dk).}
}


\maketitle
\begin{abstract}
This paper deals with two main issues regarding the short polar codes: the potential of FEC-assisted decoding and optimal code concatenation strategies under various design scenarios. Code concatenation and FEC-assisted decoding are presented systematically, assuming a packetized system. It is shown that FEC-assisted decoding can improve frame error rate of successive cancellation decoding arbitrarily, at the expense of some coding rate loss and decoding complexity linearly increasing with the number of codewords in the frame. This is compared with list decoding whose complexity grows linearly with the list size as well as the number of codewords. Thereafter, the frame construction procedure and decoding algorithm are developed in a realistic framework. Taking into consideration the effective throughput of the transmission protocol, the problem of optimal design of concatenated codes is formulated under polar code length, frame length and target frame-success-rate constraints. Simulations are performed assuming both additive white Gaussian noise and Rayleigh fading channels. It is shown that the divide-concatenate strategy for long frames does not lead to any considerable gain. It is also shown that the performance of FEC-assisted decoding of frames is improved as the frame length increases while the conventional successive cancellation decoding undergoes a dramatic performance loss.
\end{abstract}

\IEEEpeerreviewmaketitle

\section{Introduction}

\IEEEPARstart{P}{olar} codes \cite{5075875} have attracted notable interest for their ability to provably achieve the symmetric capacity of discrete memoryless channels with low complexity decoding as the code length tends to infinity. The originally proposed successive cancellation (SC) decoding could attain this performance with computational complexity $O(n\log n)$, with $n$ being the code length. Computational complexity and latency are two major bottlenecks when dealing with long codes. Hence, shorter codes are practically of interest, especially for low complexity and resource constrained applications e.g. massive machine-to-machine communication or wireless sensor networks, or latency sensitive applications e.g. mission critical machine-type of communication for steering and control. Moreover, in most of such applications only short packets are used. However, SC decoder performs poorly for short codes since the polarization effect cannot be fully utilized when the code length is not long enough. This has lead to invention of several encoding and decoding algorithms for polar codes that require a slightly higher complexity and rate compromise in exchange for improved performance. 

It is shown that concatenated with other codes such as Reed-Solomon (RS) \cite{5513508} and \cite{6816518}, LDPC \cite{6891165}, BCH and convolutional \cite{7028538} codes, short polar codes can perform much better. Additionally, it was noticed that systematic polar codes have a significant advantage in terms of bit error rate compared to the original non-systematic codes \cite{5934670}. List decoding \cite{7055304} was proposed to improve the performance of the SC decoder with complexity $O(Ln\log n)$ where $L$ is the list size. The idea is to keep a list of the recent $L$ most likely decoding paths while moving on the decoding graph instead of moving only on a single path. Only the codeword with the highest probability is selected at the end. Although it can approach the maximum likelihood performance, it was still behind the state-of-the-art LDPC and Turbo codes \cite{7055304}. Nonetheless, cyclic redundancy check (CRC)-assisted systematic polar coding with SC-list decoding performs similarly, or can even outperform these codes, provided that a sufficiently large list size is used for decoding\cite{7055304}. In this approach originally presented in \cite{6297420}, a CRC checksum is used to find the actual codeword in the list instead of selecting the most likely one.  

A typical physical layer (PHY) frame consists of multiple codewords. For successful delivery of the packet, all should be decoded correctly. Hence, the computational complexity of list decoding of $N_{CW}$ codewords with list decoding would be $O(N_{\text{cw}}Ln\log n )$ at best, i.e. when the efficient and numerically stable implementation in \cite{7055304} is used. We consider the idea originally presented in \cite{6279525} and later in \cite{6816518} for RS-polar concatenated codes to relax the requirement of a list for each codeword and instead deploy multiple SC decoders in parallel when a frame of multiple polar codewords is decoded. There are several advantages in doing so. Most importantly, the total required number of operations can be reduced down to $O(N_{\text{cw}}n\log n )$. In addition to the parallelized design, one can benefit from more capable forward error correction (FEC) codes instead of the CRC error detection codes to achieve this objective. In \cite{6816518} the code parameters are derived to achieve a target frame error rate. However, the tradeoffs between the code lengths as well as the rates of the concatenated codes for a given frame length have not been addressed. Frame length is often designed according to several different factors such as maximum latency, sampling rate, hardware limitations, multiple access and collision probabilities, etc. in realistic applications and is not arbitrarily selected to achieve a target frame error rate. Obviously, the polarization effect can be utilized more effectively in longer polar codes. It is not yet clear whether e.g. a low-rate, long polar code combined with a high rate short FEC code can perform better or the opposite combination, i.e., when a strong FEC code is used with less polarized codewords. We will address these tradeoffs in this paper by taking into consideration the effective throughput of the transmission protocol in use. Since the concatenation of the polar codes only rises for short codes (long polar codes can already achieve the channel capacity), we focus only on short code lengths (i.e. not more than several hundred bits).  
\begin{figure*}[!t]
\centering
\includegraphics[width=5in]{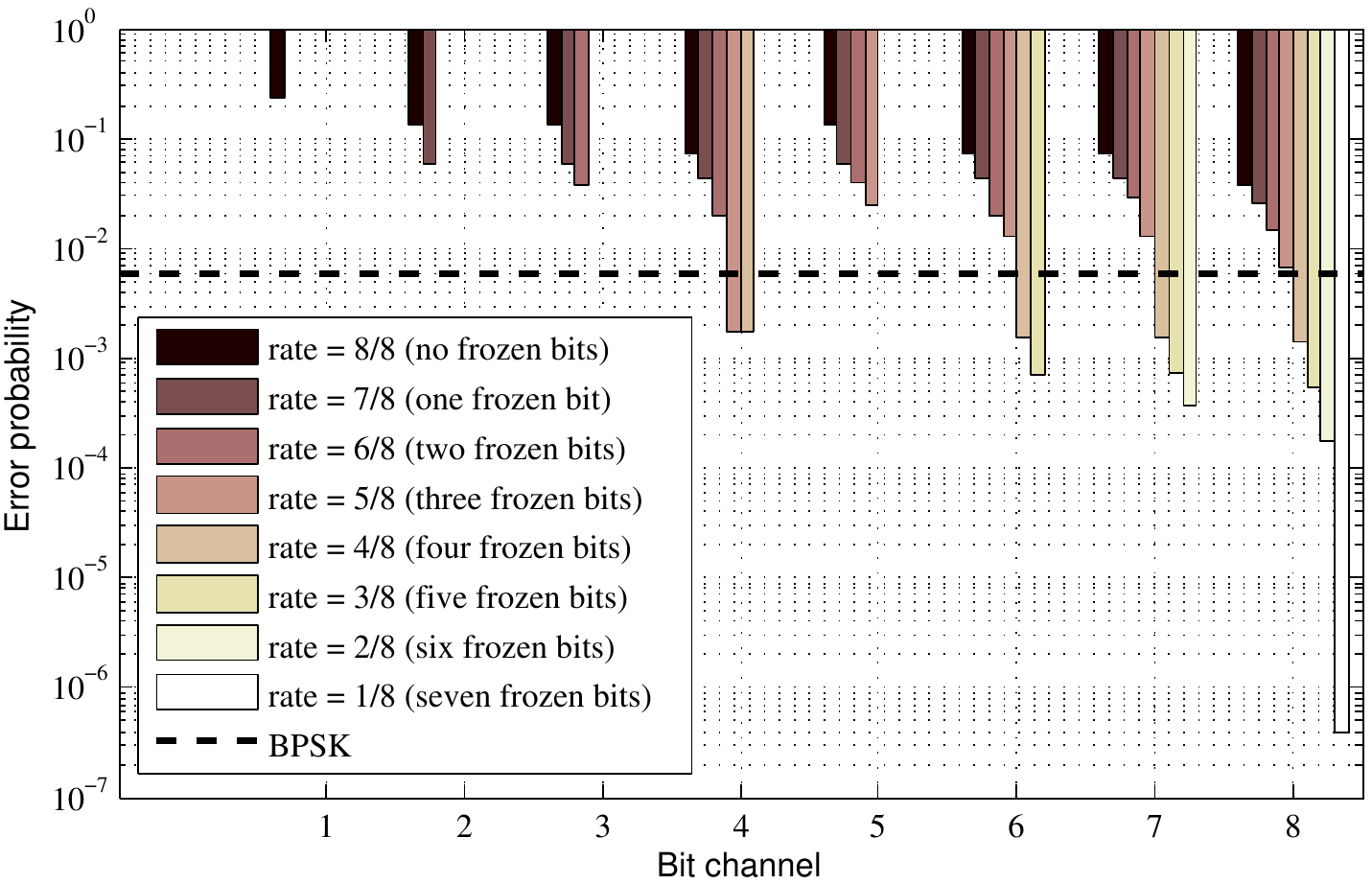} 
\caption{The error probabilities of different bit channels in a length-8 polar code of various rates based on Monte Carlo simulations at $\text{SNR}=5$ dB.}
\label{fig_ber8}
\end{figure*}
\section{Background}
This section provides the preliminaries of polar codes and the background on code concatenation.
\subsection{Polar Codes}  
Polar codes are the first linear block codes that can provably achieve the symmetric capacity of discrete memoryless channels. Consistent with the original definition of the polar codes \cite {5075875}, for a polar code of length $n$ assume the generator matrix
\begin{equation}
{\bf G}_n={\bf B}_n{\bf F}^{\otimes i},
\label{eq_pol_enc}
\end{equation}
where ${\bf B}_n$ is a permutation matrix known as the bit-reversal, ${\bf F}\triangleq \begin{bmatrix} 1 & 1\\  1 & 0 \end{bmatrix}$ is the polarization kernel, $i=\log_2(n)$, and ${\bf F}^{\otimes i}$ is the $i$'th Kronecker power of ${\bf F}$.
Assume the column vector ${\bf u}=[u_1,...\,,u_n]^\top$ of length $n$, where ``$^\top$'' is the vector transpose symbol. To construct a polar code of rate $r=k/n$, first a subset of size $n-k$ of indices in ${\bf u}$ denoted by ${\mathbb{F}}$ and referred to as``frozen'' bits are set to predefined values known for the decoder. These values are denoted by $u_{\mathbb{F}}$. The remaining $k$ elements  referred to as the ``information'' bits belong to the set ${\mathbb{A}}$ which is called the information set. In this way, $k$ message bits can be substituted in the bit positions identified in ${\mathbb{A}}$. The corresponding codeword $\mathbf{c}=[c_1,...\,,c_n]^\top$ can be given by
\begin{equation}
{\bf c}={\bf G}_n^\top{\bf u}.
\label{eq_pc}
\end{equation}
An interesting feature of the polar codes is that different elements of the information set have different levels of reliability. More explicitly, every element in the information set $\mathbb{A}$ can be seen as a channel with a unique capacity. Hence, each one is often called a separate \emph{bit channel}. Due to the polarization effect, some bit channels are more \emph{degraded}, i.e. the capacity of those channels is less and some bit channels become upgraded, i.e.  their capacity improves. The frozen bits are usually selected from the most degraded bit channels while the information bits are mapped to most upgraded channels.
For simplicity, we define a polar code  $\mathcal{C}_{\text{p}}$ of rate $r_{\text{p}}=k_{\text{p}}/n_{\text{p}}$ with a six-tuple $(n_{\text{p}},k_{\text{p}},\mathbb{F},u_{\mathbb{F}},\mathbf{w}, \boldsymbol{\varepsilon})$, where $n_{\text{p}}$ and $k_{\text{p}}$ are the code and message lengths, $\mathbf{w}=(w_1,...\,,w_{n_{\text{p}}})$ is the vector of bit channel indices sorted from the most to the least degraded, and $\boldsymbol{\varepsilon}=(\varepsilon_1,...\,,\varepsilon_{n_{\text{p}}})$ is the vector of error probabilities corresponding to each bit channel. To illustrate this, the error probabilities of different bit channels of a length-8 polar code are calculated by simulation and are plotted in Fig.\ref{fig_ber8} for various code rates and signal-to-noise-ratio (SNR) equal to 5 dB. The dashed line represents the bit error probability corresponding to hard-decision decoding of a binary phase shift keying (BPSK) system. Note that all bit channels are not available at every rate. Hence, the number of bars are different across the bit channels (for instance, the first bit channel is only available at rate $r_p=8/8$). Consider the polar code of rate $r_p=8/8$ (i.e. the one with no frozen bits). It can be seen that the most and least degraded bit channels are the bit channels ``1'' and ``8'' respectively. However, none of the bit channels can be considered upgraded in comparison with a single use of the channel with uncoded transmission (i.e. the BPSK system). This is due to the propagation of error in the decoder to the next bits after a false decoding of the preceding bits. When the code rate increases, the degraded bits are omitted from the decoding procedure since they are categorized as the frozen bits. Therefore, the probability of error will reduce. Also note that the order of the bit channels according to their degradation is $\mathbf{w}=(1, 2, 3, 5, 4, 6, 7, 8)$. It can be observed that only the reliability of the bit channels of code rates below and equal to 1/2 is better than that of uncoded transmission.

Decoding of the polar codes is successively performed on a decoding graph such that when an information bit is decoded, it can disentangle the dependencies in a way that the next bits can be decoded with the knowledge of the decoded bit. Hence, it is called successive cancellation decoder. A high-level presentation of SC decoding of a desired bit in a polar codeword is given in Algorithm \ref{alg_sc_decode}. Assume a codeword $\mathbf{c}$ of length $n$ is transmitted using binary antipodal signaling through an additive white Gaussian noise (AWGN) channel with noise variance $\sigma^2$. 
At the receiver side, denote the received signal by $\mathbf{y}=(y_1,... , y_{n})$. The relationship between $\mathbf{y}$ and $\mathbf{u}$ can be seen through a synthesized vector channel with transition probabilities denoted by $W_n$. When $b-1$ bits are decoded, the $b$'th bit is decoded by looking at the log-likelihood ratio that considers the available knowledge (the decoded bits $(\hat{u}_{w_1},... , \hat{u}_{w_{b-1}})$ and the vector $\mathbf{y}$)
\begin{equation}
l^{[b]}_n=\log\left(\frac{W_n^{[b]}\left((y_1, ... , y_n),(\hat{u}_{w_1},... , \hat{u}_{w_{b-1}})|0\right)}{W_n^{[b]}\left((y_1,... , y_n),(\hat{u}_{w_1},... , \hat{u}_{w_{b-1}})|1\right)}\right).
\end{equation}
Calculation of the above likelihood ratio when arbitrary codes are used is not straightforward. However, the Kronecker power in (\ref{eq_pol_enc}) and the specific selection of the polar code kernel make this procedure simple by providing a recursive procedure given by \cite[Eq (75) and (76)]{5075875} that only relies on the bit-level likelihood ratio
\begin{equation}
l^{[b]}_1=\frac{2}{\sigma^2}y_b,\, \forall b \in \{1,\,... ,\, n\}.
\end{equation}

\begin{algorithm}
    \SetKwInOut{Input}{Input}
    \SetKwInOut{Output}{Output}
    \Input{received signal vector $\mathbf{y}$, frozen set $\mathbb{F}$, frozen values $u_{\mathbb{F}}$}
    \Output{reconstructed data bit $\hat{u}_b$}
\uIf {$b \in \mathbb{F}$}
{
Set $\hat{u}_b$ equal to the frozen value of $u_b$
}
\uElse
{
Calculate $l^{[\mu]}_b$\\
\uIf{$l^{[\mu]}_b>0$}
{Set $\hat{u}_b=u_b$}
\uElse
{Set $\hat{u}_b=1$}
}
    \caption{SC-decode($\mathbf{y},\mathbb{F}, u_{\mathbb{F}},b$): decoding of $u_b$}
\label{alg_sc_decode}
\end{algorithm}

Although the decoding complexity is only $O(n\log n)$\cite{5075875}, the performance of SC decoder for short and moderate codes is not promising. Therefore, many improved decoding schemes are proposed. Most notable is the SC-list decoder \cite{7055304} with complexity $O(L n\log n)$ that can approach performance of the maximum likelihood decoder. A SC-list decoder is essentially a SC-decoder that keeps track of the best $L$ decoding paths on the decoding graph and keeps the list updated with the best new $L$ paths after each hard decision is made. At the end, only the most likely path is selected.

Exact calculation of the error probabilities of various polar bit channels is computationally prohibitive due to the exponential complexity of density evolution. Instead, there are efficient schemes that can estimate these probabilities with less complexity. Most notable is the Gaussian approximation (GA) which is based on the fact that the log-likelihood ratio of the received bits are normally distributed. This approach is proposed in \cite{6279525} and \cite{6823688}, and is used in this paper to calculate the error probabilities of various bit channels. Namely, the average value of the log-likelihood ratio can first be approximated by the following recursion \cite{6279525}
\begin{align}
\E\left(l^{[b]}_n\right)=
\begin{cases} 
\phi^{-1}\left(1-{\left(\phi \left( \E(l^{[(b+1)/2]}_{n/2}) \right)\right)}^2 \right), & \text{if } b \text{ is odd} \\ 
2\E(l^{[b/2]}_{n/2}), & \text{if } b \text{ is even},
  \end{cases}
\end{align}
in which $\E()$ denotes expectation and $\phi(x)$ can be fitted with the exponential model \cite{910580}
\begin{equation}
\phi(x)=
  \begin{cases} 
\exp\left({0.4527x^{0.86}+0.0218}\right), & \text{if }  0<x\le10\\ 
\frac{\sqrt{\pi}}{2}\exp\left(-\frac{x}{4}\right) \left(1-\frac{10}{7x}\right),& \text{if }x>10,
  \end{cases}
\end{equation}
and the termination rule is $\E(l^{[1]}_{1})=2/{\sigma^2}$. The approximated bit error probability corresponding to the $b$'th bit channel conditioned on correct decoding of the previous bits is given by
\begin{equation}
\varepsilon_{b}=Q\left(\sqrt{\frac{\E(l^{[b]}_{n})}{2}} \right).
\end{equation}
A comparison between the simulated error probabilities for short polar codes and the corresponding GA-based probabilities can be found in the Appendix.

\subsection{Concatenated Codes}
The idea of concatenating two different codes was first introduced and analyzed by Forney in his doctoral thesis \cite{forney1965concatenated}. It was a method to yield the performance of long codes by breaking the decoding complexity into segments of shorter concatenated codes. He proved that in this way the probability of error vanishes exponentially while the decoding complexity grows algebraically with the length of the shorter segments. However, it is required that the code length of the inner code be logarithmic with the length of the outer code which can considerably reduce the effectiveness of polarization when the polar codes are used as the inner code. Concatenation of polar codes has shown its potential to approach the channel capacity with low complexity and several schemes have been proposed. Another important difference between the presented FEC assisted decoding of polar codes and the conventional decoding of concatenated codes is that the decoding of the inner code is performed sequentially, where one of the outer codewords are decoded in each step. In the conventional approach, all inner codewords are decoded first to come up with the outer codewords. 

\begin{figure*}[!t]
\centering
\includegraphics[width=5in]{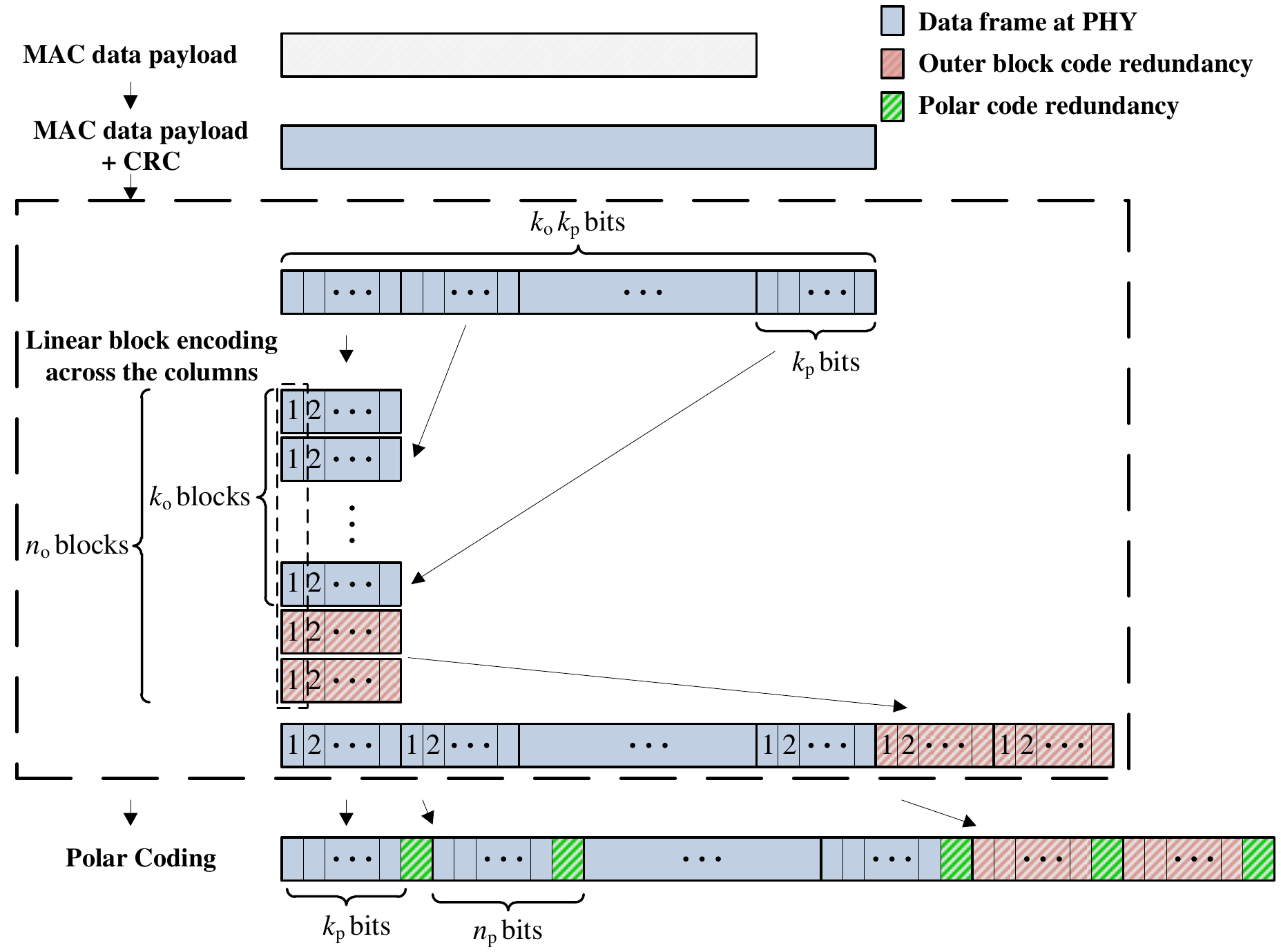} 
\caption{The encoding procedure ($\beta=1$). MAC data payload is encoded with a CRC code at the MAC layer. At the physical layer, the frame is divided to $n_{\text{o}}$ blocks of length $k_{\text{p}}$ and the block elements corresponding to different positions across all the blocks are used for liner block coding. Finally, each block is independently encoded with a ($n_{\text{p}}$,$k_{\text{p}}$) polar code.}
\label{fig_enc}
\end{figure*}

\section{Concatenation of Short Polar Codes}
We consider a point-to-point packet-based communication system. The frame construction at the The medium access control (MAC) and PHY layer is depicted in Fig.~\ref{fig_enc}. Data of interest is first encoded by an error detection code at the MAC  layer and then the frame along with the checksums undergoes a concatenated encoding at the physical layer. Essentially, a $(n_{\text{o}},k_{\text{o}},t_{\text{o}})$ linear block code (LBC) $\mathcal{C}_o$ is exploited for FEC-assisted decoding of the inner polar code $\mathcal{C}_{\text{p}}$, where $n_{\text{o}}$, $k_{\text{o}}$, and $t_{\text{o}}$ are the code and message lengths (in bits) and the correction capability of the outer LBC, respectively. Instead of using the whole frame for one concatenation, one can also devide the frame to a number of super-segments (for instance $\beta$) and perform this procedure on each super-segment just like in Fig.~\ref{fig_enc}. Note that different bit channels of the polar code have different levels of reliability and hence require different levels of protections. Although we can allocate  different rates of the outer code for each bit channel, at the end we need to pad the remaining bits with zeros to fill the empty segments which will lead to the same number of redundant bits of the fixed allocation. Therefore, we assume outer codes with equal rates for all polar bit channels to fully utilize the available frame. The frame construction procedure is presented in Algorithm \ref{alg_construction}.

\begin{algorithm}
    \SetKwInOut{Input}{Input}
    \SetKwInOut{Output}{Output}
    \Input{Data frame $\mathbf{\tilde{m}}$, $\beta$, Code parameters of $\mathcal{C}_{\textnormal{p}}$, $\mathcal{C}_{\textnormal{o}}$}
    \Output{Encoded frame $\mathbf{\tilde{p}}$}
Divide the frame $\mathbf{\tilde{m}}$ to $\beta$ super-segments $\mathbf{\check{m}}$\\
\For {{\bf all } \textnormal{super-segments} $\mathbf{\check{m}}$}
{
$\alpha=\lceil\frac{L_{\textnormal{MAC}}}{k_{\textnormal{p}}}\rceil$\\
Pad $\alpha k_{\textnormal{p}}-L_{\textnormal{MAC}}$ zero bits to the end of the frame\\
 \For{$v=1,\,...,\, k_{\textnormal{p}}$}
      {	
		{\tt //separate the bits in a column}\\	
	  	$\mathbf{m}_v=\mathbf{\check{m}}(z,\,z+k_{\textnormal{p}},\,...,\,z+\alpha k_{\textnormal{p}})$\\
		{\tt //encode using the outer code}\\	
		$\mathbf{c}_v=\text{encode (} \mathbf{m}_v,\, \mathcal{C}_{\textnormal{o}} \text{)}$\\
		{\tt //build the encoded frame}\\	
		$\mathbf{\check{p}}(v,\,v+n_{\textnormal{p}},\,...,\,v+ n_{\textnormal{o}} n_{\textnormal{p}})=\mathbf{c}_v(1,\,... ,\, n_{\textnormal{o}})$\\
		}					
\For{$z=1,\,...,\, n_{\textnormal{o}}$}
		{
		{\tt //encode with the polar code}\\	
		$\mathbf{\check{p}}((z-1)n_{\textnormal{p}}+1,\,...,\, z n_{\textnormal{p}})=\text{encode (} \mathbf{\check{p}}((z-1)n_{\textnormal{p}}+1,\,...,\, (z-1)n_{\textnormal{p}}+k_{\textnormal{p}}),\, \mathcal{C}_{\textnormal{p}} \text{)}$\\
 		} 
}    
Merge all encoded super-segments $\mathbf{\check{p}}$ to form the encoded frame $\mathbf{\tilde{p}}$
    \caption{Frame construction by concatenating the polar code $\mathcal{C}_{\textnormal{p}}(n_{\textnormal{p}},k_{\textnormal{p}})$ and the outer code $\mathcal{C}_{\textnormal{o}}(n_{\textnormal{o}},k_{\textnormal{o}})$.}
\label{alg_construction}
\end{algorithm}

Due to several advantages of BCH codes such as simple decoding\footnote{Syndrome decoding is a minimum distance decoding that exploits linearity to reduce decoding complexity using a reduced lookup table.} and the precise control over their correction capability of multiple bits, we focus on binary BCH codes for the outer LBC. It is shown that the decoding complexity of the binary BCH codes can be reduced to $O(t\sqrt{n_{\textnormal{o}}})$ \cite{6034253}. The CRC code is traditionally used to verify the integrity of the frame after decoding at the receiver side. The frames with failed checksums are discarded and should be retransmitted. 

\begin{figure*}[!t]
\centering
\includegraphics[width=5in]{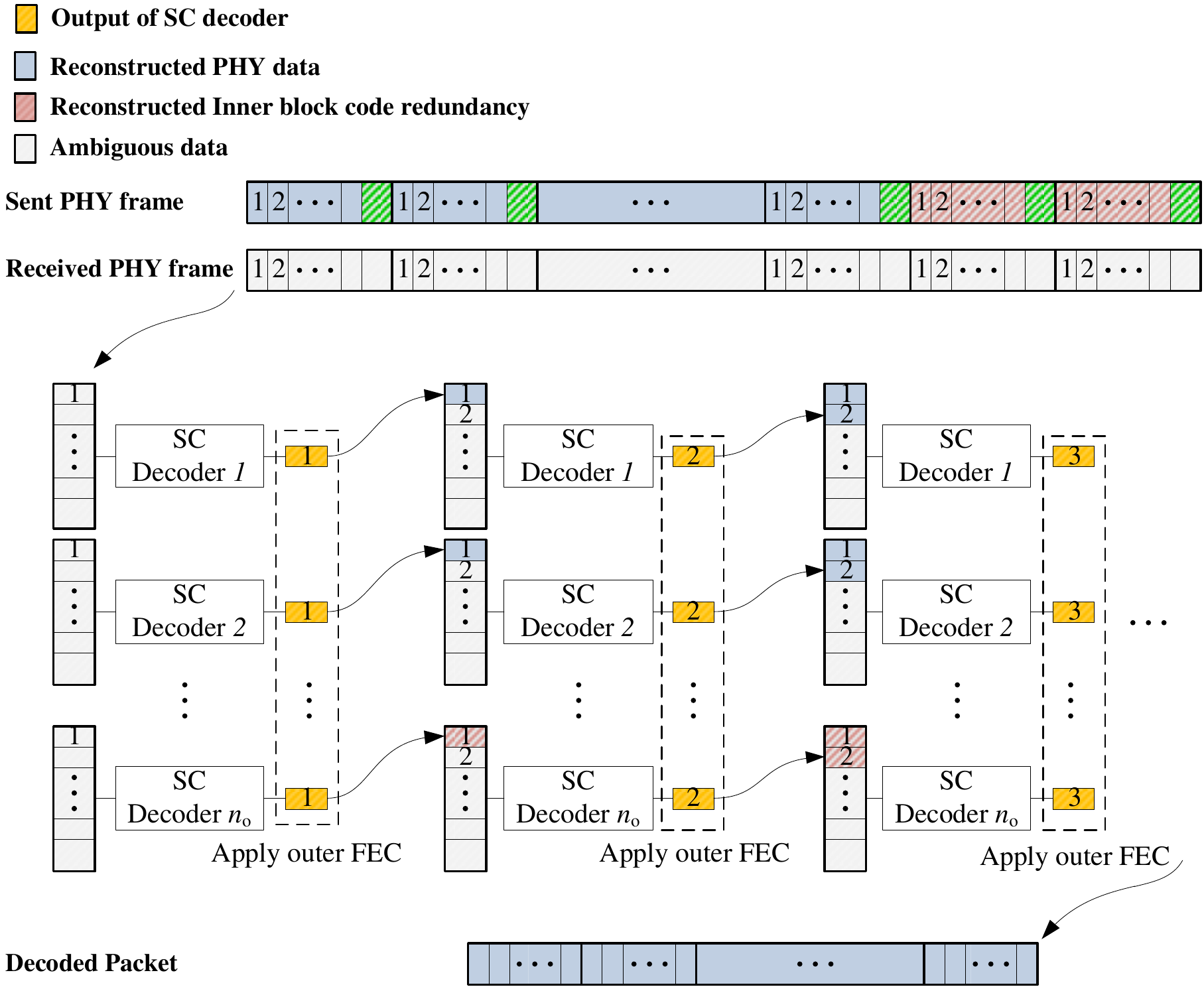} 
\caption{FEC assisted decoding of the polar codewords ($\beta=1$).}
\label{fig_dec}
\end{figure*}

\begin{algorithm}
    \SetKwInOut{Input}{Input}
    \SetKwInOut{Output}{Output}
    \Input{Received signal frame $\mathbf{\tilde{y}}$, $\beta$, Code parameters of $\mathcal{C}_{\textnormal{p}}$, $\mathcal{C}_{\textnormal{o}}$}
    \Output{Data frame $\mathbf{\hat{m}}$}
Divide $\mathbf{\tilde{y}}$ to $\beta$ super-segments $\mathbf{\check{y}}$\\
\For {{\bf all } \textnormal{super-segments} $\mathbf{\check{y}}$}
{
Initialize $\mathbf{y}_1,\,... ,\, \mathbf{y}_{n_{\textnormal{o}}}$ by dividing $\mathbf{\check{y}}$ to $n_{\textnormal{o}}$ vectors corresponding to $\mathbf{c}_1,\,... ,\, \mathbf{c}_{n_{\textnormal{o}}}$\\
Initialize $\mathbf{\check{m}}(1,\,... ,\, k_{\textnormal{p}}n_{\textnormal{o}})=0$\\
 \For{$v=1,\,...,\, k_{\textnormal{p}}$}
      {		
		\For{$z=1,\,...,\, n_{\textnormal{o}}$}
			{		
{\tt //decode the $v$'th bit of the $z$'th polar codeword}\\	
$\mathbf{c}_v(z)=\text{sc-decode} (\mathbf{y}_z,\, \mathbb{F}_z,\,u_{\mathbb{F}_z},\,w_{n_{\textnormal{p}}-k_{\textnormal{p}}+v})$\\	  
}
{\tt //apply FEC on the outer code}\\	
$\mathbf{c}_v=\text{correct} (\mathbf{c}_v,\mathcal{C}_{\textnormal{o}})$\\
{\tt //distribute the corrected bits in the reconstructed frame}\\	
$\mathbf{\check{m}}(v,\,v+k_{\textnormal{p}},\,...,\,v+(n_{\textnormal{o}}-1) k_{\textnormal{p}})=\mathbf{c}_v(1,\,... ,\, n_{\textnormal{o}})$
{\tt //treat the decoded value as a new frozen bit in each polar codeword}\\	
\For{$z=1,\,...,\, n_{\textnormal{o}}$}
{
{\tt //add the correct bit index to the frozen set}\\	
$\mathbb{F}_z=\mathbb{F}_z+w_{n_{\textnormal{p}}-k_{\textnormal{p}}+v}$\\
{\tt //add the corrected value to the frozen values}\\	
$u_{\mathbb{F}_z}(w_{n_{\textnormal{p}}-k_{\textnormal{p}}+v})=\mathbf{c}_v(z)$
}
		}					
$\mathbf{\hat{m}}=\mathbf{\check{m}}(1,\,... ,\, k_{\textnormal{p}}k_{\textnormal{o}})$
}    
Merge all decoded super-segments $\mathbf{\hat{m}}$ to form the frame $\mathbf{\tilde{m}}$
    \caption{FEC-assisted decoding of the polar code $\mathcal{C}_{\textnormal{p}}(n_{\textnormal{p}},k_{\textnormal{p}})$ using the outer code $\mathcal{C}_{\textnormal{o}}(n_{\textnormal{o}},k_{\textnormal{o}})$ (sequential design).}
\label{alg_sec_dec}
\end{algorithm}

\subsection{FEC-Assisted Successive Cancellation Decoding}
The key fact to consider is that in any level of decoding if a faulty hard decision is made by the SC decoder, it can propagate to the next levels and with a high probability the decoded message include several bit errors. Therefore, it is very important to first make sure the decisions made by the SC decoder are correct and then proceed to the next levels. This is the basis for the FEC-assisted decoding where an outer block code can be used to protect different bit levels against false decisions. In the traditional decoding of the concatenated codes the inner code is first completely decoded and then the outer decoder uses the result to reconstruct the message. However, in the FEC-assisted decoding procedure illustrated in Fig.~\ref{fig_dec}, the decoding of the inner polar code is not a one-time procedure but a successive one. At first, only the first information bit is decoded. However, the first information bit of \emph{all} the codewords are decoded in \emph{parallel}. After this point, all the parallel SC decoders halt and pass the decoded column to the outer decoder for FEC correction. The corresponding bits of the corrected outer codeword are then passed again back to each of the SC decoders and now they all can proceed to decode the second information bit. This procedure continues until all information bits are decoded. The computational complexity of this approach is the number of the outer codewords times the complexity of the SC decoder (plus the complexity of the outer code). A sequential implementation of the FEC-assisted decoding is given in Algorithm \ref{alg_sec_dec}.

\subsection{Single Concatenation}
An important question in designing the concatenated codes is how to optimally design the code parameters of the outer and the inner code. We first assume single concatenation which means the frame is not divided into smaller segments before concatenation. Identically, in this section we look for the best combinations of the codes, regardless of the frame length. We denote the total number of concatenations in the frame by an integer $\beta$. Fig.~\ref{fig_enc} represents the case with single concatenation $\beta=1$. When $\beta>1$, the frame can be divided into $\beta$ super-segments where each segment can again be represented similarly to Fig.~\ref{fig_enc}. We will discuss the case with $\beta>1$ in the next section and consider $\beta=1$ here. Assume a polar code $\mathcal{C}_{\text{p}}$ of length $n^*_{\text{p}}$ is given. Consider the set
\begin{equation}
\mathbb{S}_{\text{o}}=\big\{\mathcal{C}^{(j)}_{\text{o}} \colon n^{(j)}_{\text{o}} = 2^j-1\colon \forall j \le n_{\text{o,max}} \big\},
\label{eq_so_np}
\end{equation}
of BCH codes of length $n^{(j)}_{\text{o}}$ over integer $j$ values that can be concatenated, where $n_{\text{o,max}}$ is the highest possible length of the outer code. Note that the total number of parallel SC decoders is given by the length of the outer code. Hence, it is practically required to assume a maximum value for $n_{\text{o}}$.  In order to find the best combination, we focus on the effective throughput of the transmission protocol defined as
\begin{equation}
T(\gamma) =  r_{\text{o}}r_{\text{p}}{\mathsf{FSR}}(\gamma),
\label{eq_trp}
\end{equation}
where $r_{\text{o}} = k_{\text{o}}/n_{\text{o}}$, $r_{\text{p}}=k_{\text{p}}/n_{\text{p}}$, and $\mathsf{FSR}(\gamma)$ is the frame success rate in a given SNR denoted by $\gamma$. $\mathsf{FSR}(\gamma)$ is calculated by averaging the number of correctly delivered frames to the total sent frames. An optimization problem can be formulated as follows
\begin{equation}
\begin{aligned}
& \underset{\mathcal{C}_{\text{o}},k_{\textnormal{p}}}{\text{maximize}}
& & T(\gamma) \\
& \text{subject to}
& & n_{\text{p}}=n^*_{\text{p}} .
\end{aligned}
\label{eq_t_np}
\end{equation}
The outer code can be found using Algorithm \ref{alg_con_np}. Algorithm \ref{alg_findoptimal} considers the error probabilities of different bit channels of the polar code to find the optimal codes $\mathcal{C}^{*}_{\text{o}}$ and $\mathcal{C}^{*}_{\textnormal{p}}$. First, consider the auxiliary probability 
\begin{equation} 
P_{\text{x}}(\epsilon)=\sum_{z=0}^{t_{\text{o}}}\begin{pmatrix}n_{\text{o}}\\ z\end{pmatrix}\epsilon^z(1-\epsilon)^{(n_{\text{o}}-z)},
\end{equation}
which is the probability of occurrence of at most $t_{\text{o}}$ bits of errors in a column of length $n_{\text{o}}$ bits when the bit error probability is $\epsilon$. For a SC decoder concatenated with an outer LBC with correction capability $t_{\text{o}}$ and code length $n_{\text{o}}$ the probability of successful decoding is 
\begin{equation}
\mathsf{FSR}(\gamma)={\left(\prod_{b\in\mathbb{A}}{P_\textnormal{x}\left(\varepsilon_b(\gamma)\right)}\right)}^{\beta},
\label{eq_fsr}
\end{equation}
since when there is no more than $t_{\text{o}}$ errors in each column, they can all be corrected by the LBC. For simplicity, a lower bound 
\begin{equation}
\mathsf{FSR}(\gamma)\ge{\left[P_\textnormal{x}(\varepsilon_{w_{q}}(\gamma))\right]}^{\beta k_{\textnormal{p}}},
\label{eq_pb}
\end{equation}
can also be derived by assuming the error probabilities of all bit channels are equal to that of the least reliable bit channel $w_q$, where $q=n_{\textnormal{p}}-k_{\textnormal{p}}+1$. Note that the FSR of the original SC decoder can be given by
\begin{equation}
\mathsf{FSR}(\gamma)={\left(P_{\text{cw}}\right)}^{N_{\text{cw}}},
\end{equation}
where
\begin{equation}
P_{\text{cw}}=\prod_{b\in\mathbb{A}}{\left(1-\varepsilon_b\right)},
\end{equation}
is the success probability of each codeword and $N_{\text{cw}}=n_{\text{o}}$ in our setup. It would be intuitive to know the behavior of the FSR across different code lengths. In Fig. \ref{fig_fsr}, the FSR of polar codes with different code lengths versus the code rates are compared assuming a $(7, 4)$ outer code at $\textnormal{SNR}=5$ dB. When the code rate is increased the FSR approaches zero. We are interested in the threshold at which the FSR falls as the code rate increases from zero to one. The granularity of the rates across various code lengths is different. Explicitly, fewer possible code rates exist for shorter codes. A sharp breakdown with a distinguished ``threshold'' code rate is the FSR characteristics of long codes while shorter codes have graceful FSR.
\begin{figure}[!t]
\centering
\includegraphics[width=3in]{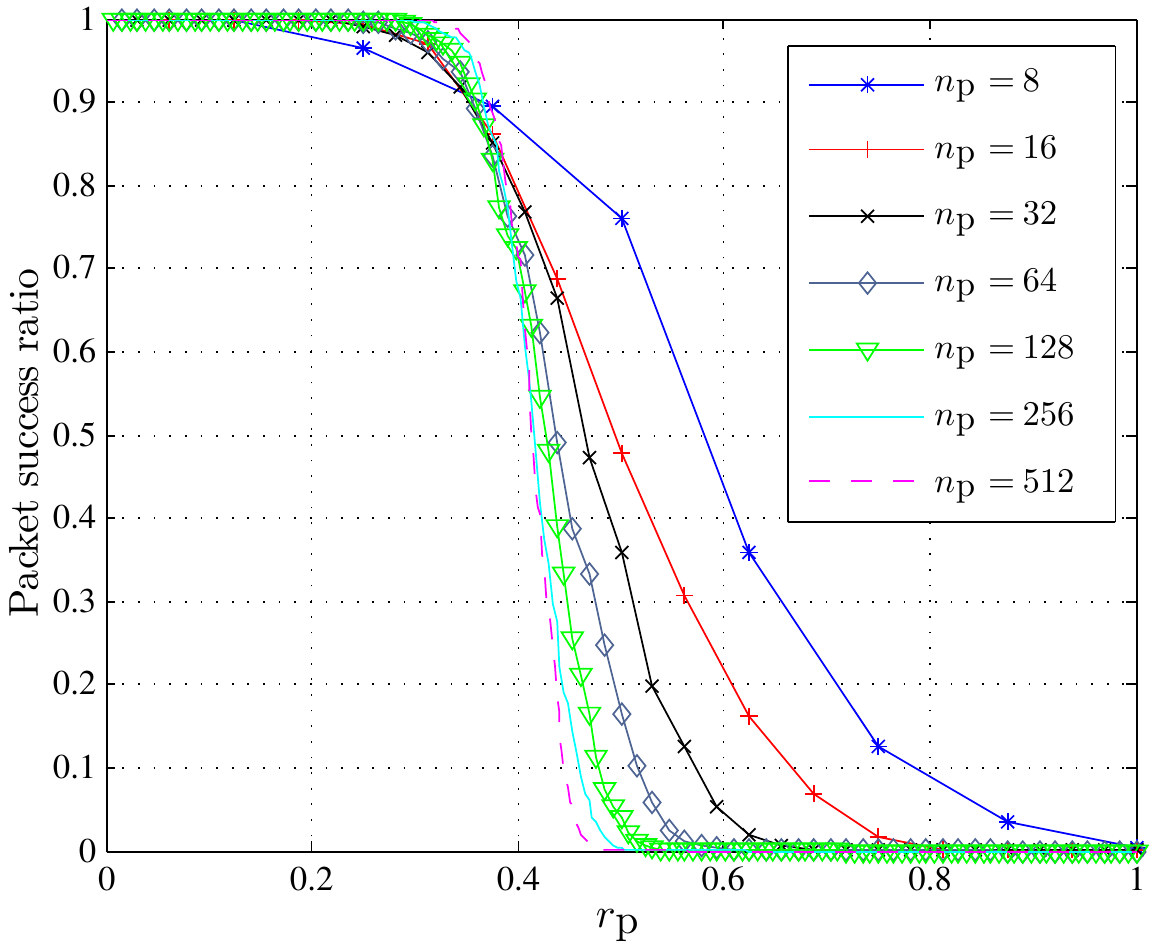} \\
\caption{Frame success rate for various polar code lengths with a $(7,4)$ outer code at $\text{SNR}=5$ dB. Possible code rates are limited in shorter codes which leads to a lower granularity.}
\label{fig_fading}
\end{figure}
{
\begin{algorithm}
    \SetKwInOut{Input}{Input}
    \SetKwInOut{Output}{Output}
    \Input{$n^*_{\text{p}}, n_{\text{o,max}}$}
    \Output{$\mathcal{C}^{*}_{\text{o}},k^*_{\textnormal{p}}$}
		Initialize $\mathbb{S}_{\text{o}}$ according to (\ref{eq_so_np})\\
$(k^*_{\textnormal{p}},\mathcal{C}^*_{\textnormal{o}})={\tt FindOptimal}(n_{\textnormal{p}}, \mathbb{S}_{\textnormal{o}})$\\

    \caption{Optimal concatenated code design for the constrained polar code length scenario: single concatenation ($\beta=1$).}
\label{alg_con_np}
\end{algorithm}}

\begin{algorithm}
    \SetKwInOut{Input}{Input}
    \SetKwInOut{Output}{Output}
        \Input{$n^*_{\textnormal{p}}, \mathbb{S}_{\textnormal{o}}$, $[L_{\textnormal{PHY}}]$ (optional)}
    \Output{$k^*_{\textnormal{p}}, \mathcal{C}^*_{\textnormal{o}}$}
$T^*=0$\\
\For{{\bf all }$\mathcal{C}^{(j)}_{\textnormal{o}}\in \mathbb{S}_{\textnormal{o}}$}
		{
			\For{$t_{\text{o}}=1,...\,,2^{j-2}-1$}
			{
			If exists, find the valid code and the corresponding $k_{\text{o}}$ in $\mathcal{C}^{(j)}_{\textnormal{o}}(n^{(j)}_{\text{o}},?,t_{\text{o}})$ code\\
			
\uIf{$L_{\textnormal{PHY}}$ \textnormal{ is constrained}}
	{		
	$\beta= \lfloor \frac{L_{\text{PHY}}}{n_{\textnormal{p}}n_{\textnormal{o}}}\rfloor$\\
}
\Else
{
	$\beta=1$

}
Initialize $\mathbb{F}=\varnothing$ (the empty set)\\

				\For{$b=1,...\,, n_{\textnormal{p}}$}
				{
				$\mathbb{F}=\mathbb{F}+u_{w_b}$\\									
				$k_{\text{p}}=n_{\text{p}}-b+1$\\
				$\epsilon=\varepsilon_{w_b}( \gamma_{\textnormal{t}})$ and\\
				Calculate $\mathsf{FSR}$ as in (\ref{eq_fsr}) or (\ref{eq_pb}) \\
						
				Calculate $T$\\
					\uIf  {$T>T^*$}{
					$T^*=T$\\
					$\mathcal{C}^{*}_{\text{o}}=\mathcal{C}^{(j)}_{\textnormal{o}}$\\
$u_{\mathbb{F}}=0$\\					$k^{*}_{\textnormal{p}}=k_{\textnormal{p}}$\\					
					}					
				}
			}	
		} 
				
 \caption{{\tt FindOptimal}: Find the optimal  combination of the polar - outer code}	
\label{alg_findoptimal}	
\end {algorithm}

\subsection{Constrained Frame Length}
The second design case study we consider is when the PHY layer frame length $L_{\text{PHY}}$ is constrained. This situation can occur for example due to multiple access considerations when each user needs to communicate in a limited timeslot. As it can be inferred from Fig.~\ref{fig_enc}, for a given length $L_{\text{PHY}}$ the final design should satisfy $\beta n_{\text{o}}n_{\text{p}} \le L_{\text{PHY}}$, where $\beta$ is an integer number representing the number of possible super-segments. The equality is for the ideal case that no bit padding is required for the combination of the LBC and the polar code. It follows that $L_{\text{MAC}}=\beta k_{\text{o}}k_{\text{p}}$ bits can be delivered in this way. In the first step, we group all possible polar codes $\mathcal{C}_{\text{p}}$ in the set
\begin{equation}
\mathbb{S}_{\text{p}}=\bigg\{\mathcal{C}^{(i)}_{\text{p}} \colon {n^{(i)}_{\text{p}} = 2^i\colon \forall i \le\left \lfloor \log_2{\frac{L_{\text{PHY}}}{n_{\text{o,min}}}} \right \rfloor \bigg\}},
\label{eq_sp}
\end{equation}
where ${n_{\text{o,min}}}$ is the smallest possible valid length of the outer code, $i$ values are integer, and $\left \lfloor\cdot \right \rfloor$ represents the floor operation. ${n_{\text{o,min}}}=7$ in case of BCH codes. 

For each polar code $\mathcal{C}^{(i)}_{\text{p}}$ in $\mathbb{S}_{\text{p}}$ there exists a set
\begin{equation}
\mathbb{S}^{(i)}_{\text{o}}=\bigg\{\mathcal{C}^{(j)}_{\text{o}} \colon n^{(j)}_{\text{o}} = 2^j-1\colon \forall j \le\left \lfloor \log_2{\frac{L_{\text{PHY}}}{n^{(i)}_{\text{p}}}} \right \rfloor \bigg\},
\label{eq_so}
\end{equation}
of BCH codes of length $n^{(j)}_{\text{o}}$ over integer $j$ values that can be concatenated. The constrained PHY layer frame length scenario can be formulated as the following optimization problem
\begin{equation}
\begin{aligned}
& \underset{\beta,\mathcal{C}_{\text{o}},\mathcal{C}_{\textnormal{p}}}{\text{maximize}}
& & T=\frac{\beta k_{\text{o}}k_{\text{p}}\mathsf{FSR}(\gamma)}{L_{\text{PHY}}} \\
& \text{subject to}
& & \beta n_{\text{o}}n_{\text{p}} \le L_{\text{PHY}}.
\end{aligned}
\label{eq_optimization}
\end{equation}
Similar to the fixed code length scenario, Algorithm \ref{alg_con_phy} can be used to find the optimal combination. 
{
\begin{algorithm}
    \SetKwInOut{Input}{Input}
    \SetKwInOut{Output}{Output}
    \Input{$L_{{\textnormal{PHY}}}, n_{\text{o,min}}$}
    \Output{$\mathcal{C}^{*}_{\text{o}},\mathcal{C}^{*}_{\textnormal{p}}$}
$T^*=0$\\
Initialize $\mathbb{S}_{\textnormal{p}}$ according to (\ref{eq_sp})\\
 \For{{\bf all }$\mathcal{C}^{(i)}_{\textnormal{p}}\in \mathbb{S}_{\textnormal{p}}$}
      {
		Initialize $\mathbb{S}^{(i)}_{\text{o}}$ according to (\ref{eq_so})\\
				$\mathcal{C}_{\textnormal{o}}={\tt FindOptimal}(n_{\textnormal{p}}, \mathbb{S}_{\textnormal{o}},L_\textnormal{PHY})$\\

}
    \caption{Optimal concatenated code design for constrained PHY frame length scenario: multiple concatenations ($\beta\ge1$).}
\label{alg_con_phy}
\end{algorithm}}

A similar algorithm can also be used when the length of the frame $L_{\text{MAC}}$ is constrained. This scenario can happen for instance when a certain frame length should support a given data rate with the available buffer size at the higher layers. Again, by referring to Fig.~\ref{fig_enc} it follows that the two concatenated codes should satisfy $\beta k_{\text{o}}k_{\text{p}}\ge L_{\text{MAC}}$. The optimization problem in (\ref{eq_optimization}) is changed accordingly as follows
\begin{equation}
\begin{aligned}
& \underset{\beta,\mathcal{C}_{\text{o}},\mathcal{C}_{\textnormal{p}}}{\text{maximize}}
& & T=\frac{ L_{\text{MAC}}\mathsf{FSR}(\gamma)}{\beta n_{\text{o}}n_{\text{p}}} \\
& \text{subject to}
& & \beta k_{\text{o}}k_{\text{p}} \ge L_{\text{MAC}}.
\end{aligned}
\label{eq_optimization_2}
\end{equation}
\subsection{Unconstrained Design with a Target FSR}
 In the design phase usually a pair of target ($\gamma_{\text{t}}, \mathsf{FSR}_{\text{t}})$ is known and it is desired to achieve a $\mathsf{FSR}\ge\mathsf{FSR}_{\text{t}}$ for the received SNRs equal to or higher than $\gamma_{\text{t}}$. 
A pair of codes $\mathcal{C}_{\text{p}}$ and $\mathcal{C}_{\text{o}}$ can be concatenated and achieve the target ($\gamma_{\text{t}}, \mathsf{FSR}_{\text{t}})$ if
\begin{equation}
{\left(\prod_{b\in\mathbb{A}}{P_\textnormal{x}\left(\varepsilon_b(\gamma)\right)}\right)}^{\beta}\ge\mathsf{FSR}_{\text{t}}.
\label{eq_px_mac}
\end{equation}
The optimal codes that result in the minimum frame length can be found by solving the optimization problem 
\begin{equation}
\begin{aligned}
& \underset{\mathcal{C}_{\text{o}},\mathcal{C}_{\textnormal{p}}}{\text{minimize}}
& & \beta n_{\textnormal{p}}n_{\textnormal{o}} \\
& \text{subject to}
& & \mathsf{FSR}\ge\mathsf{FSR}_{\text{t}}, \\
& & & \beta k_{\textnormal{p}}k_{\textnormal{o}}\ge L_{\text{MAC}}.
\end{aligned}
\end{equation}
The search procedure is similar to the constrained MAC frame length scenario, except the objective should change appropriately.
\subsection{Optimal Codes for Fading Channels}
In case the concatenated codes are designed for fading channels, the first step would be to discretize the channel state space to a sufficiently large number of states. Assume 
\begin{equation}
f_{\Gamma}(\gamma) = \frac{1}{\bar{\gamma}}e^{-\gamma/\bar{\gamma}},
\end{equation}
is the probability density function of the fading, where $\bar{\gamma}$ is the average SNR, and there are a total number of $S$ channel states identified by the boundary SNR values $\{\gamma_{1},...\,,\gamma_{S}\}$. A channel state is denoted by $s$ where $s\in\{1,...\,, S\},$ can be identified by an average SNR $\bar{\gamma}_{s}=\int_{\gamma_{s-1}}^{\gamma_{s}}\gamma f_{\Gamma}(\gamma)d\gamma$ as well as a state probability $\Pr\{s\}=\int_{\gamma_{s-1}}^{\gamma_{s}}f_{\Gamma}(\gamma)d\gamma$. Obviously, we can also calculate the average value for the effective throughput in each channel state. In the limit when $S \to \infty$, the instantaneous values can be used as the channel states. However, the error characteristics of the codes may not be available for all instantaneous SNR values. The exact value of the average effective throughput of each channel state denoted by $\bar{T}_s$ should be calculated by integrating (\ref{eq_trp}) over the corresponding SNR interval with the fading pdf. We however, approximate it by substituting  (\ref{eq_trp}) with the average SNR of the corresponding channel state for simplicity. The effective throughput of the transmission protocol in the fading channel can then be given by
\begin{equation}
\tilde{T} = \sum_{s=1}^{S}{\bar{T}_s}\Pr\{s\}.
\end{equation}
The problem of finding the optimal codes for the fading channel and under the constrained PHY layer frame length can then be written as follows
\begin{equation}
\begin{aligned}
& \underset{\mathcal{C}_{\text{o}},\mathcal{C}_{\textnormal{p}}}{\text{maximize}}
& & \tilde{T} \\
& \text{subject to}
& & n_{\text{o}}n_{\text{p}} \le L_{\text{PHY}}.
\end{aligned}
\end{equation}
Also for the fixed polar code length scenario a similar optimization problem with the constraint given in (\ref{eq_t_np}) can be formulated and solved.
\section{Evaluation and Discussion}
\begin{figure*}[!t]
\centering
\includegraphics[width=3.5in]{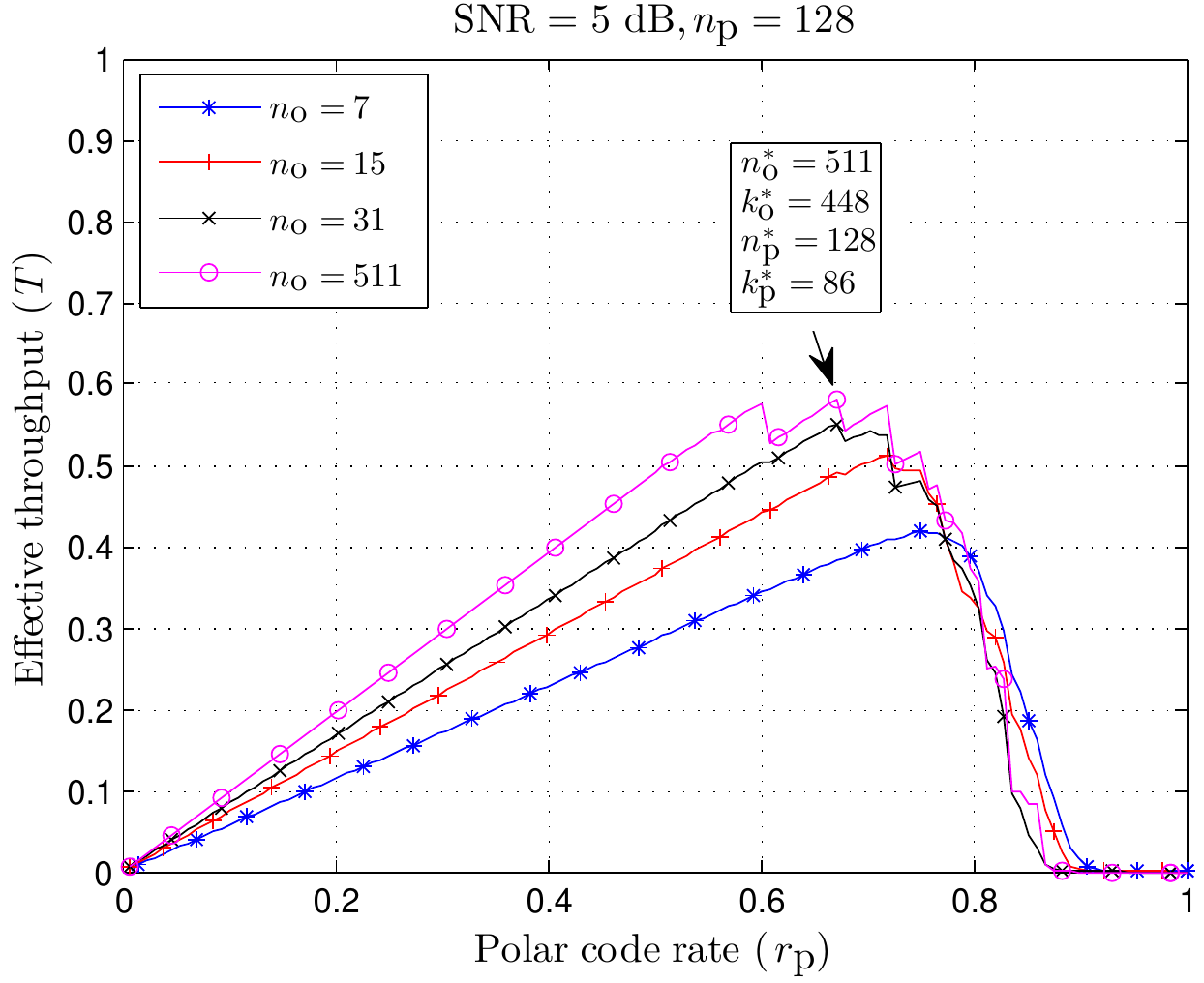} 
\includegraphics[width=3.5in]{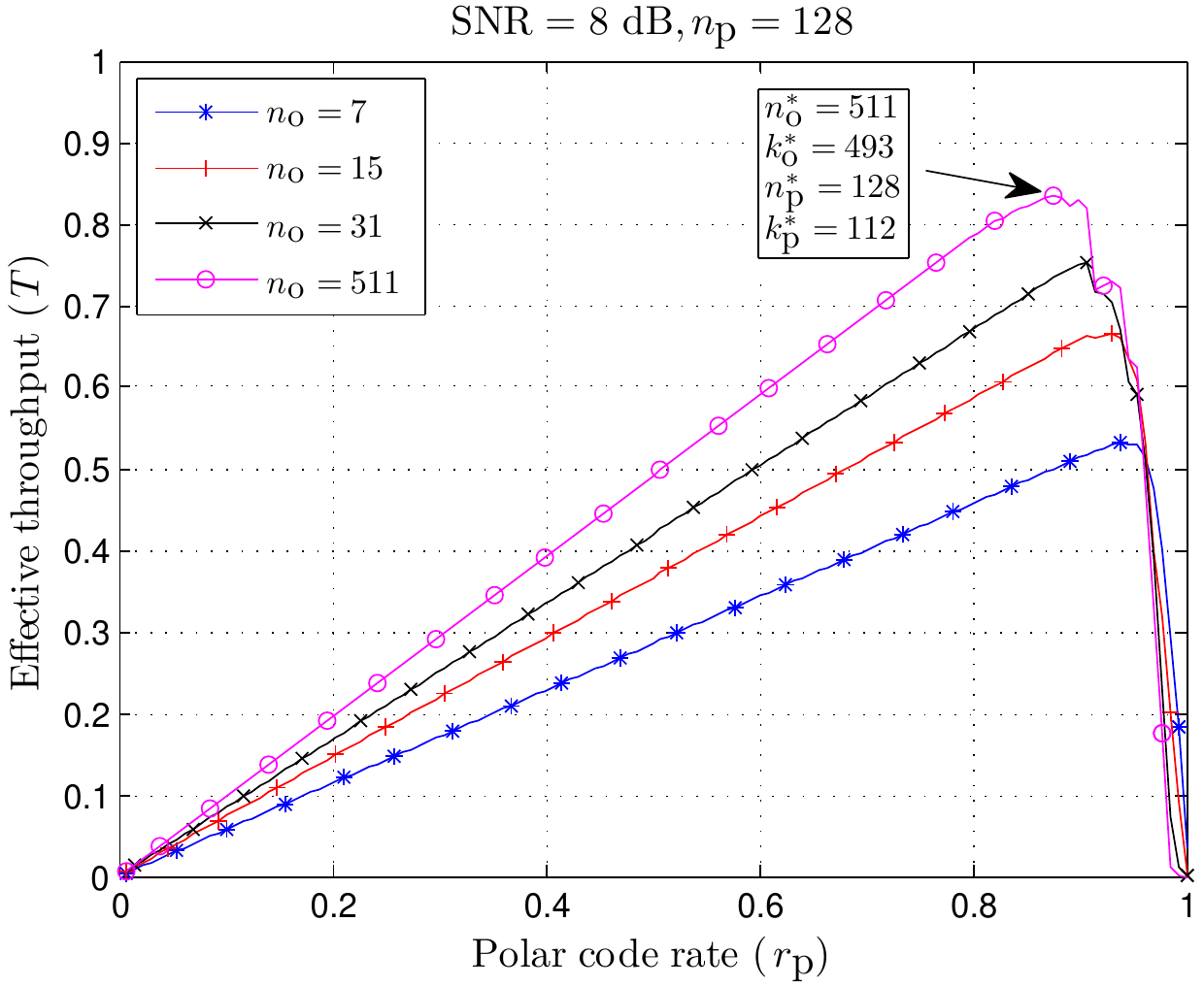} \\
\includegraphics[width=3.5in]{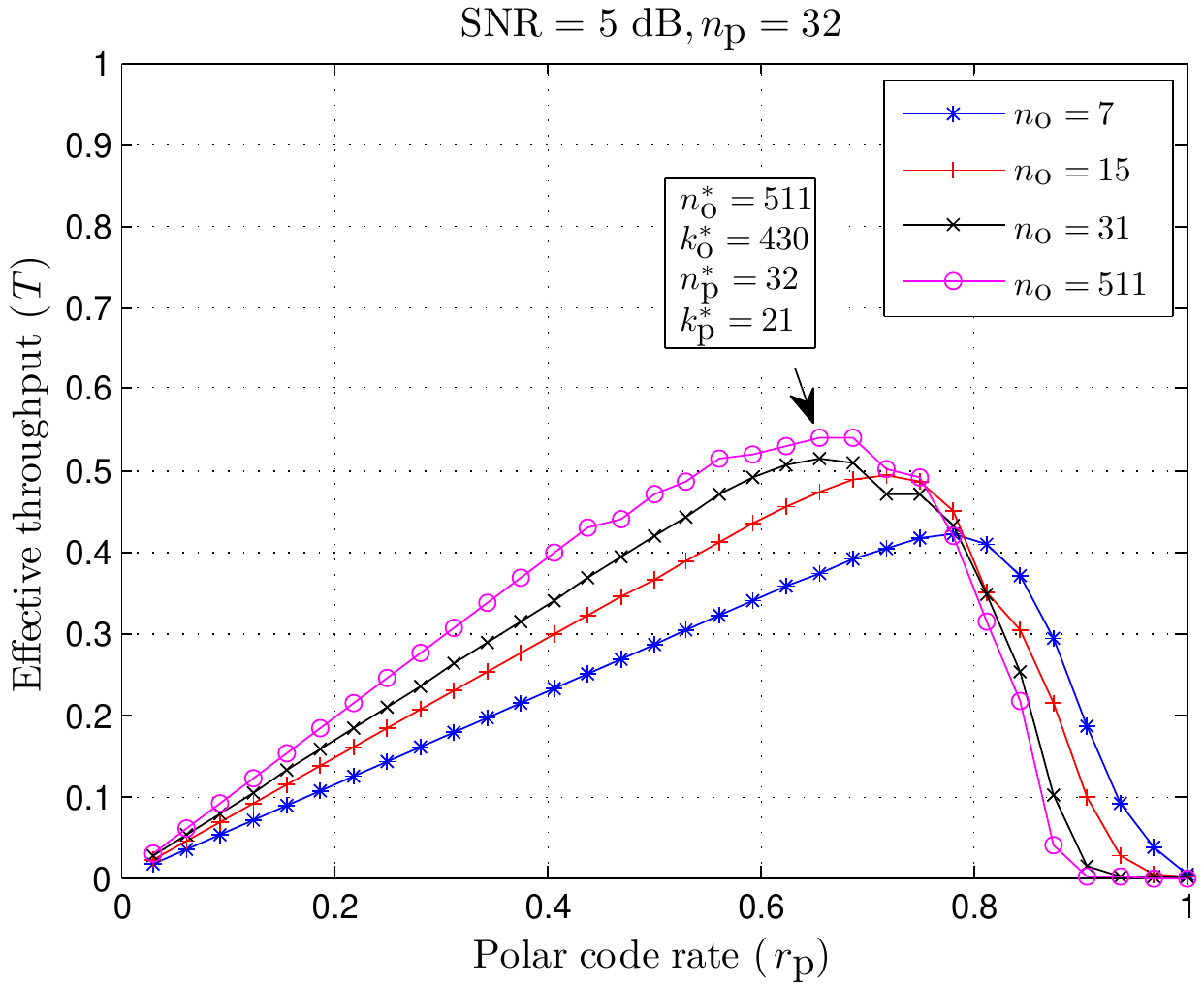} 
\includegraphics[width=3.5in]{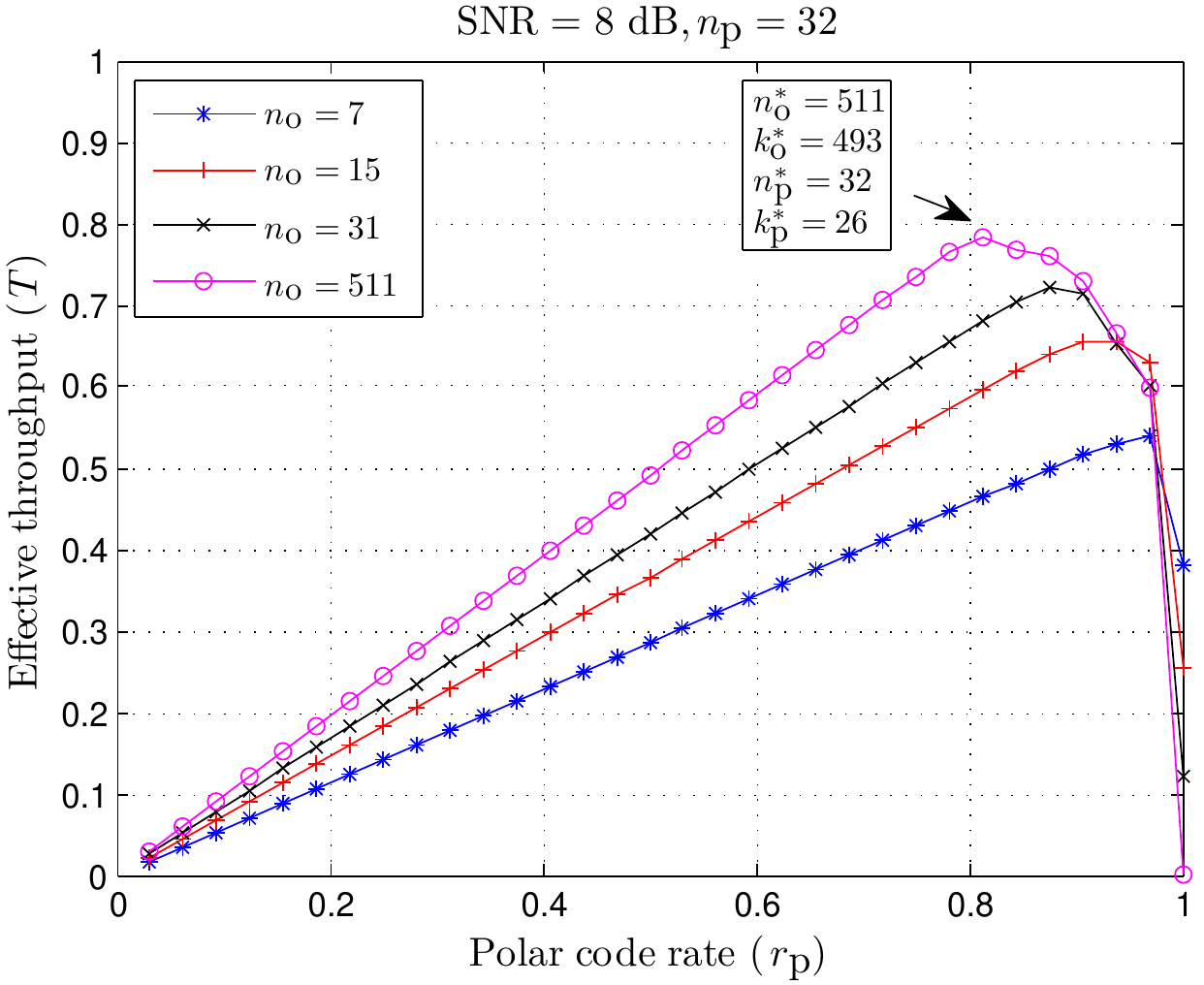} 
\caption{Optimal design of \emph{single} ($\beta=1$) concatenated codewords for fixed $n_{\text{p}}=128$ (top) and $n_{\text{p}}=32$ (bottom) at $\text{SNR}=5$ dB (left) and $\text{SNR}=8$ dB (right).}
\label{fig_fixed_np}
\end{figure*}

The concatenation of polar and the outer code can be designed based on different objectives. In the first scenario we look for the optimal design that maximizes the effective throughput with a single concatenation ($\beta=1$). We assume two different polar code lengths $n_{\textnormal{p}}=32$ and $n_{\textnormal{p}}=128$ are given and no constraint on the frame length at the physical layer is assumed. The simulation results are illustrated in Fig. \ref{fig_fixed_np}. Both (\ref{eq_fsr}) and (\ref{eq_pb}) are examined to estimate the FSR. However, only the results corresponding to the more accurate estimate (\ref{eq_fsr}) are presented. With (\ref{eq_pb}), the resulting optimal effective throughput is almost the same but the optimal rates of the outer and the polar code are slightly different. It is noted that in a given polar code length the maximum achievable throughput depends on the length of the outer code. The higher code lengths lead to higher values. However, a form of saturation can be observed in the sense that the performance improvement is not considerable when the outer code lengths greater than $n_{\text{o}}=31$ are used.

\begin{figure}[!h]
\centering
\subfigure[]{
\includegraphics[width=3in]{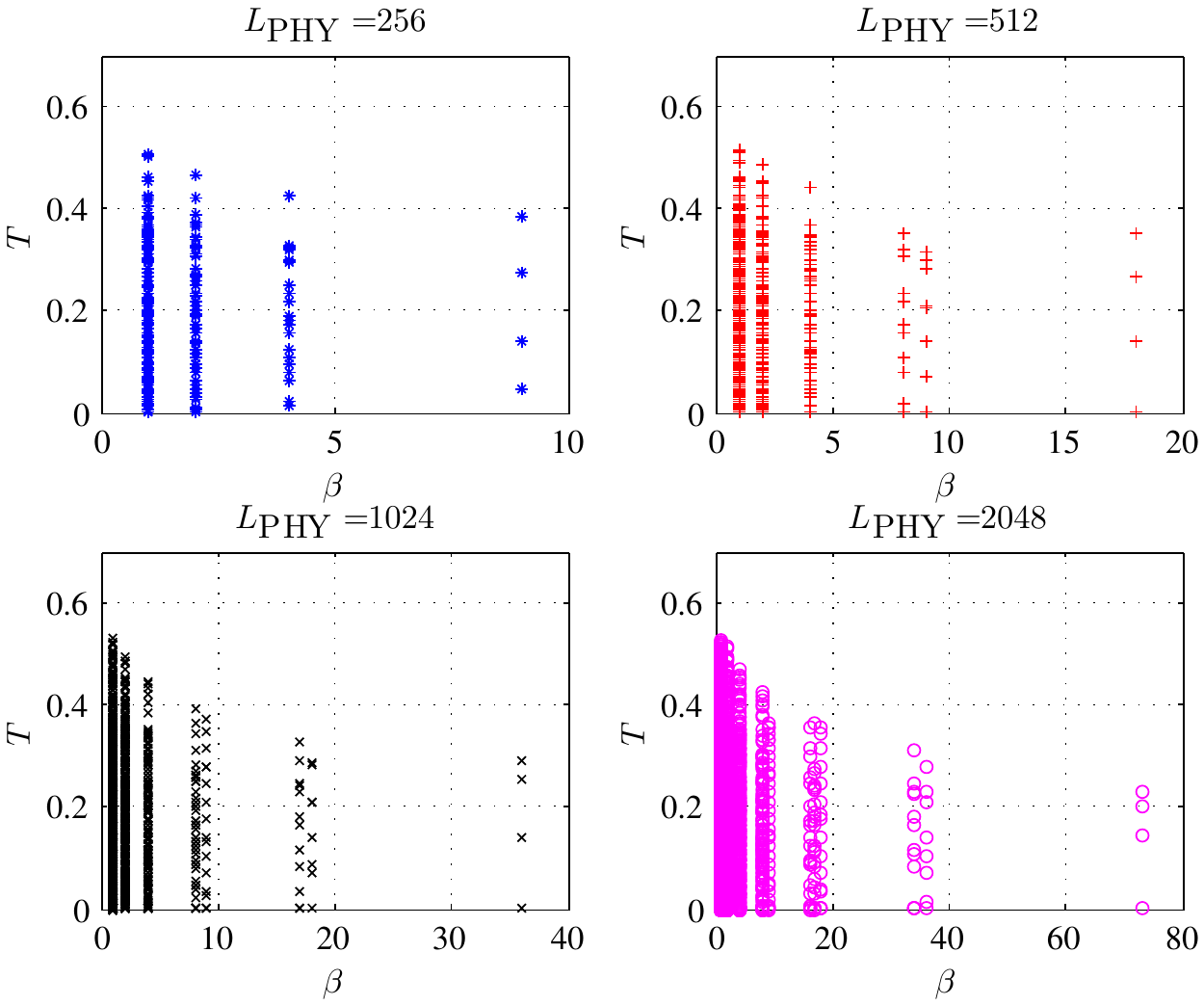}
\label{fig_phy_a}} \\
\subfigure[]{
\includegraphics[width=3in]{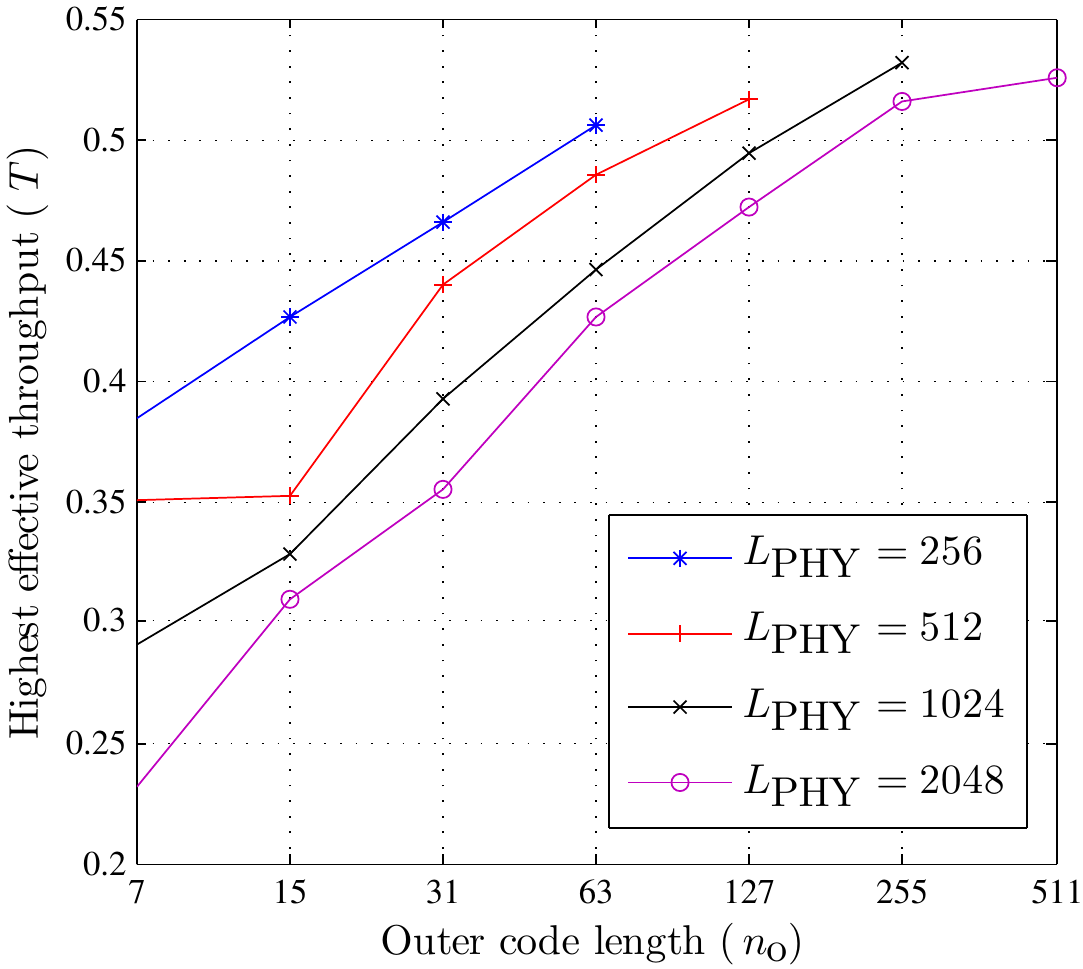}
\label{fig_phy_b}} \\
\subfigure[]{
\includegraphics[width=3in]{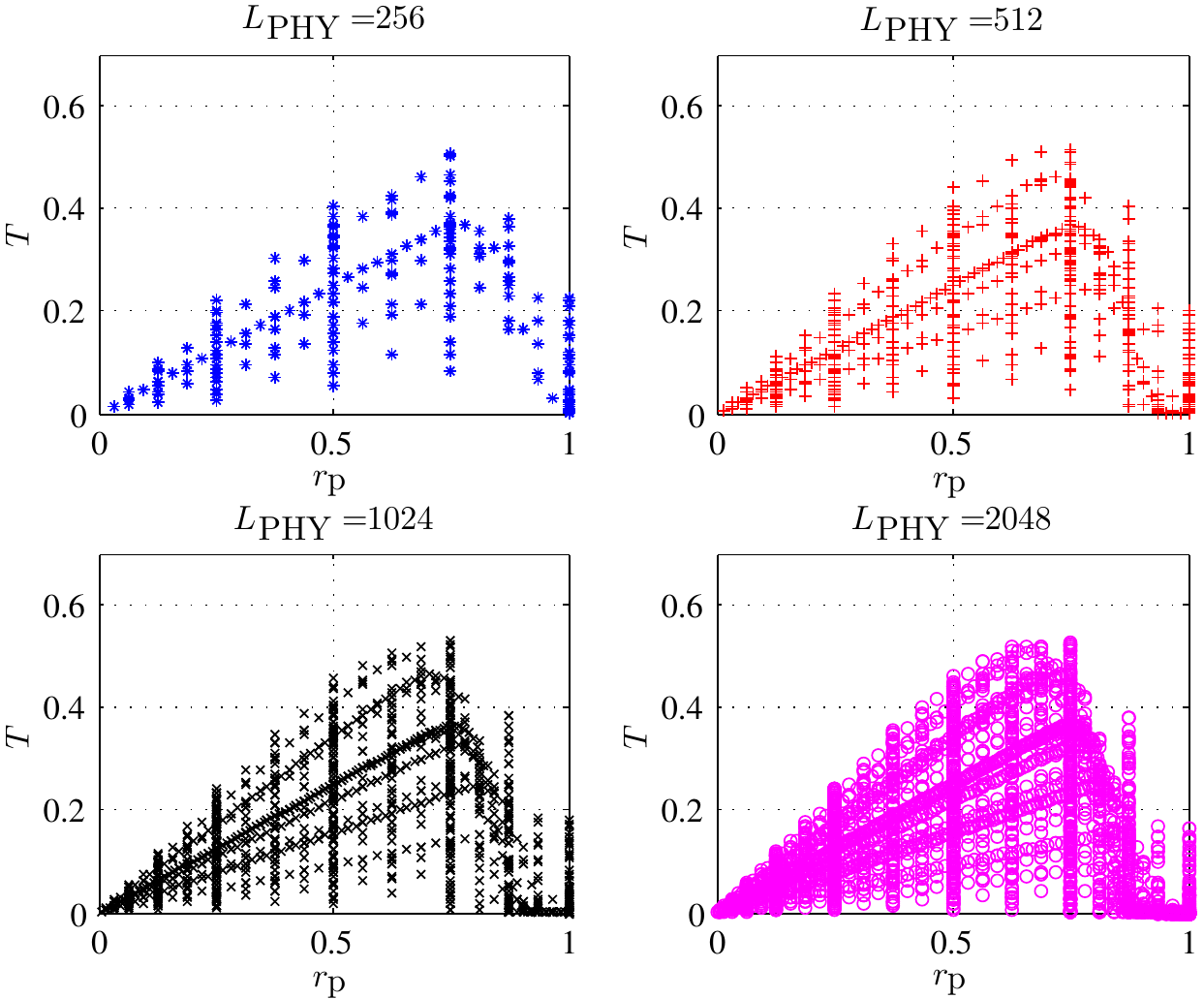}
\label{fig_phy_c}} 
\caption{Constrained PHY frame length scenario ($\text{SNR}=5$ dB). Achievable throughput versus polar code rate (a), Outer code length (b), and $\beta$ (c), respectively. }
\label{fig_phy}
\end{figure}

Now consider the scenario with a fixed PHY frame length $L_{\text{PHY}}$ where the parameter $\beta$ can be selected arbitrarily. We first observe the performance of the concatenated codes with respect to various $\beta$ and $L_{\text{PHY}}$ values in Fig. \ref{fig_phy_a}. This is an essential step to understand if the multiple concatenations can lead to any gain compared to single concatenation. It is seen that the achievable effective throughput will reduce if we divide either of the outer code or the polar code into $\beta>1$ smaller segments and repeatedly perform the concatenation for each segment. The achievable throughput versus the rate of the polar code is also depicted in Fig. \ref{fig_phy_c} and the optimal values are tabulated in Table \ref{tab_phy}. Generally speaking, the optimal throughput cannot exceed more than $53\%$, regardless of the frame length. The specific value will depend on the SNR though. Of course this limit is reached through different code combinations and $\beta$ values in various frame lengths. The only difference that the frame length can make is the total number of feasible codes which can be seen from the density of the points in the figure. The vertical lines in the figure correspond to short polar codes since the rate of the code has a coarse granularity. For instance, with $n_{\text{p}}=4$ only four different rates are possible (excluding zero). The possible combinations of the outer code contribute to different values of the effective throughput (i.e. vertical shifts) for each rate. A special case in this figure is when $r_{\text{p}}=1$, i.e. the polar code does not perform any error correction and only the correction capability of the outer code is utilized. In such cases, usually with a very small  $n_{\text{p}}$ (e.g. 4) and a large $\beta$ an effective throughput up to about $20\%$ is possible. It is improved by more than two orders of magnitude when the rate of the polar code is reduced and the polarization takes effect. However, only one frozen bit would be enough (at this specific SNR) since in all cases the $(4,3)$ polar code can achieve the highest throughput when concatenated with the appropriate outer code. A similar simulation is performed at $\text{SNR}=0$ dB and the results are given in Table \ref{tab_phy}. It can be seen that while still $\beta=1$ has the best result, more polarized bits (i.e. lower code rate with longer code lengths) are preferred at such low SNRs.

The impact of the outer code length on the effective throughput is illustrated in Fi. \ref{fig_phy_c}. A direct relationship between the length of the outer code and the maximum achievable throughput can be inferred. However, a form of saturation in the effective throughout can also be identified for $n_{\text{o}}\ge63$ at this SNR. This is actually due to the fact that the number of codewords in the frame is equal to the length of the outer code which is more elaborated in the sequel.

Similar simulations were accomplished for the constrained MAC frame length scenario which is represented in Table \ref{tab_mac}. The final optimal design corresponding to the target FSR scenario is very close to the constrained MAC frame length for practical target FSR values. However, it is not always the case. For instance, with $L_\text{MAC}=512$ a $(31,26)$ outer code with a $(32,20)$ polar code can achieve $T=0.49$ with a target FSR $0.95$ at $\text{SNR}=5$ dB using a shorter PHY frame length. Although the optimal values of $\beta$ are higher than one, other codes with $\beta=1$ can also reach very close to the same throughput. Therefore, it is concluded that under all scenarios no performance benefit (in terms of the effective throughput) can be made by dividing the frame to $\beta>1$ super-segments and applying multiple concatenations.

\begin{figure}[!h]
\centering
\includegraphics[width=3in]{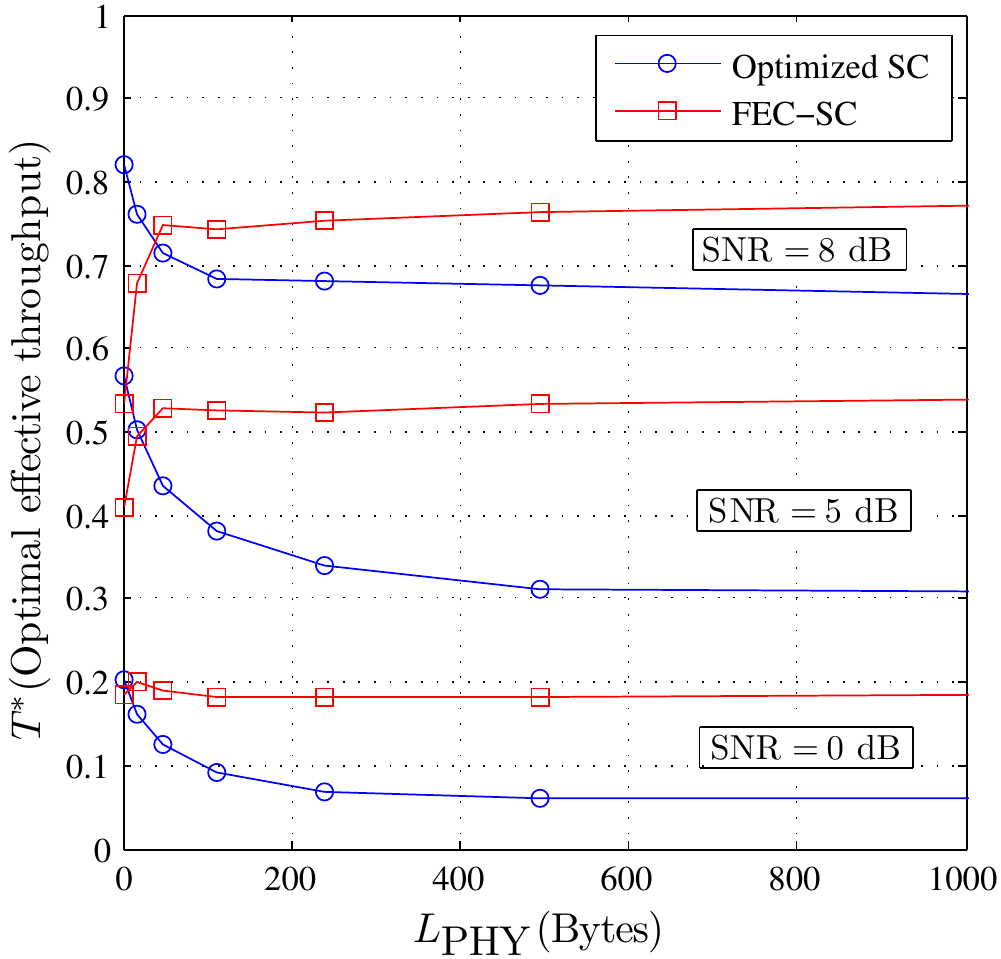} 
\caption{Maximum achievable throughput at different PHY frame lengths, assuming $n_{\textnormal{p}}=16$.}
\label{fig_t_l_phy}
\end{figure}

In the sequel we study the impact of increasing the length of the frame on the achievable throughput. In the classic systems, each frame is composed of several codewords. Longer frames are composed of more codewords. It is more likely to have one or more false decoded codewords in longer frames, hence frame error probability is higher.  The reason is that all codewords should be decoded correctly to deliver the frame successfully. In Fig \ref{fig_t_l_phy}, the performance of polar coding with SC decoder rate-optimized for each frame length is compared with that of FEC-assisted scheme, assuming $n_{\text{p}}=16$ and for various PHY frame lengths and SNRs. It is shown that the achievable throughput corresponding to the SC decoding drops with increasing the PHY frame length. The FEC-assisted scheme, however, performs exactly opposite. This is due to the joint operation of the outer and the polar code.

\begin{figure}[!h]
\centering
\includegraphics[width=3in]{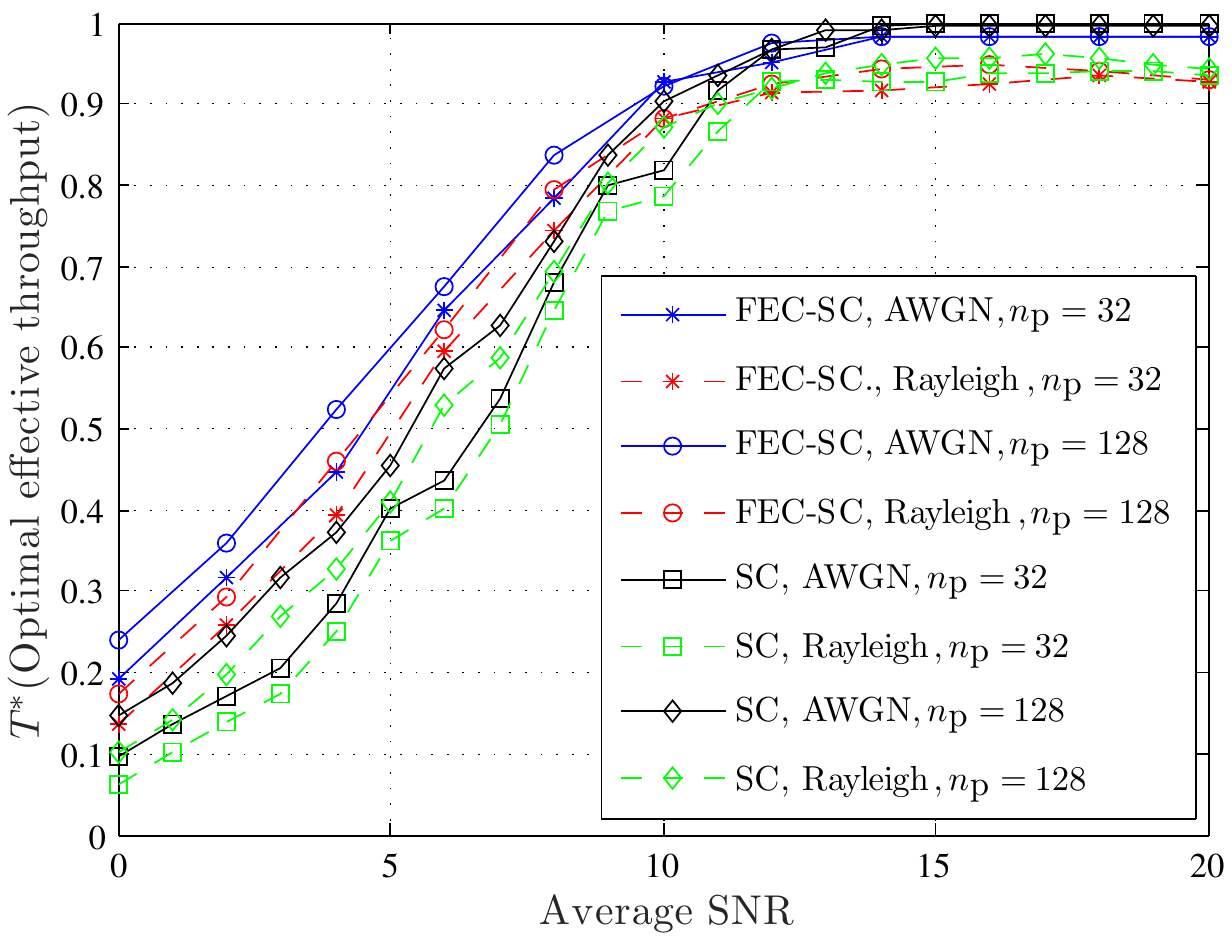} 
\caption{Maximum achievable throughput in the AWGN and the Rayleigh fading channels with $n_{\textnormal{p}}=32$ and $n_{\textnormal{p}}=128$.}
\label{fig_fading}
\end{figure}

\begin{table}[]
\centering
\caption{Optimal Values - Constrained PHY scenario}
\label{tab_phy}
\begin{tabular}{cccccc}
				            	& 		$L_{\text{PHY}}$	& 256 & 512 & 1024 & 2048 \\ \hline
\multirow{6}{*}{$5$ dB}	& $n^*_{\textnormal{p}}$ &4 & 4&4&4         \\
								& $k^*_{\textnormal{p}}$ &3 & 3&3&3         \\
								& $n^*_{\textnormal{o}} $&63 & 127&255&511         \\
								& $k^*_{\textnormal{o}} $& 45 & 92&187&376       \\
								& $\beta^*$ &1 & 1&1&1 \\
								& $T^*$  &0.51  &  0.52 &   0.53&   0.53 \\   \hline
\multirow{6}{*}{$0$ dB}	& $n^*_{\textnormal{p}}$ &16 & 16&64&128         \\
								& $k^*_{\textnormal{p}}$ &5 & 5&20&39         \\
								& $n^*_{\textnormal{o}} $&15 & 31&15&15         \\
								& $k^*_{\textnormal{o}} $& 11 & 21&11&11       \\
								& $\beta^*$ &1 & 1&1&1 \\
								& $T^*$  &0.19  &  0.18 &   0.19&   0.20  
\end{tabular}
\end{table}

\begin{table}[]
\centering
\caption{Optimal values - Constrained MAC scenario}
\label{tab_mac}
\begin{tabular}{lccccc}
				            	& 		$L_{\text{MAC}}$	& 128 & 256 & 512 & 1024 \\ \hline
\multirow{6}{*}{5 dB}	& $n^*_{\textnormal{p}}$ &4 & 16&4&64         \\
								& $k^*_{\textnormal{p}}$ &3 & 10&3&40         \\
								& $n^*_{\textnormal{o}} $&63 & 31&255&31         \\
								& $k^*_{\textnormal{o}} $& 45 & 26&171&26       \\
								& $T^*$  &0.49  &  0.50 &   0.50&   0.51 \\   
								& $\beta$ &1 & 1&1&1 \\
								& $L_{\text{PHY}}$ &252 & 496&1020&1984\\ \hline
\multirow{6}{*}{0 dB}	& $n^*_{\textnormal{p}}$ &16 & 32&64&128         \\
								& $k^*_{\textnormal{p}}$ &4 & 8&16&40         \\
								& $n^*_{\textnormal{o}} $&15 & 15&15&31         \\
								& $k^*_{\textnormal{o}} $& 11 & 11&11&26       \\
								& $\beta^*$ &3 & 3&3&1 \\
								& $T^*$  &0.16  &  0.16 &   0.17&   0.18\\
								& $L_{\text{PHY}}$ &720 & 1440&2880&3968  
\end{tabular}
\end{table}


\subsection{Fading Channel}
As it is inferred in the last section, the maximum achievable throughput depends on the channel SNR. In Fig \ref{fig_fading}, we study the optimal throughput in various channel conditions assuming AWGN and Rayleigh fading channel models, given two different lengths of the polar code $n_{\text{p}}=32$ and $n_{\text{p}}=128$ and with both SC decoder and FEC-assisted decoder. The number of codewords in each frame for the simulation of the SC decoder is $N_{cw}=511$ which is the number of codewords (as well as the length of the outer code) in the FEC-assisted scheme. Two different regimes can be identified for both channel models. The maximum achievable throughput improves constantly with SNR in the first regime. It is a consequence of the improvement in the rate of the polar code. The second regime begins at a specific SNR after which the effective throughput is saturated since the highest code rate is already achieved. Although code concatenation reduces the overall code rate by a factor of $k_{\text{o}}/n_{\text{o}}$, the throughput is improved at the medium and low SNRs compared to the SC decoder due to the correction capability of the outer code. Nonetheless at the high SNR regime, the rate reduction of the outer code takes effect. It should be noted that the optimal throughput of the SC decoder at high SNR regime is achieved indeed with no frozen bits, and is illustrated only for comparison.

\section{Conclusions}
A thorough study on polar codes concatenated with BCH linear block codes under different constraints on code and frame parameters assuming AWGN and fading channels were considered in this paper. The encoding and decoding algorithms were presented assuming a packet-based system and the advantages of FEC-assisted decoding were elaborated compared to list decoding and conventional successive cancellation decoding. Parallel deployment of FEC-assisted decoding in packet-based systems can achieve the equivalent performance of list decoding with $1/L$'th complexity where $L$ is the list size. Compared to SC decoder, the packet success ratio can arbitrarily improved by adjusting the correction capability of the outer code, which is analogous to list decoding with various list sizes. As opposed to rate-optimized SC decoder, the achievable throughput  does not degrade in FEC-assisted decoding when the length of PHY frame increase. It is concluded that dividing a long frame to shorter concatenated codes cannot lead to a higher effective throughput. Additionally, the higher the length of the outer code is, the higher is the achievable effective throughput.


\end{document}